\documentclass[12pt,a4paper]{article}
\usepackage{amssymb}
\usepackage{amsfonts}
\usepackage[onehalfspacing]{setspace}

\newtheorem{theorem}{Theorem}

\newtheorem{claim}[theorem]{Claim}

\newtheorem{definition}{Definition}

\newtheorem{proposition}{Proposition}

\newenvironment{proof}[1][Proof]{\noindent\textbf{#1.} }{\ \rule{0.5em}{0.5em}}
\input{tcilatex}
\begin{document}

\title{A Model of Competing Narratives\thanks{%
Financial support by ERC Advanced Investigator grant no. 692995 is
gratefully acknowledged. We thank Heidi Thysen, Stephane Wolton and
conference audiences at ESSET and CCET for helpful comments.}}
\author{Kfir Eliaz and Ran Spiegler\thanks{%
Eliaz: School of Economics, Tel-Aviv University and Economics Dept.,
Columbia University. E-mail: kfire@post.tau.ac.il. Spiegler: School of
Economics, Tel-Aviv University and Economics Dept., University College
London and CFM. E-mail: rani@post.tau.ac.il.}}
\maketitle

\begin{abstract}
We formalize the argument that political disagreements can be traced to a
\textquotedblleft clash of narratives\textquotedblright . Drawing on the
\textquotedblleft Bayesian Networks\textquotedblright\ literature, we model
a narrative as a causal model that maps actions into consequences, weaving a
selection of other random variables into the story. An equilibrium is
defined as a probability distribution over narrative-policy pairs that
maximizes a representative agent's anticipatory utility, capturing the idea
that public opinion favors hopeful narratives. Our equilibrium analysis
sheds light on the structure of prevailing narratives, the variables they
involve, the policies they sustain and their contribution to political
polarization.\bigskip \bigskip \bigskip \pagebreak
\end{abstract}

\section{Introduction}

It has become commonplace to claim that political disagreements can be
traced to a \textquotedblleft \textit{clash of narratives}\textquotedblright
. Going beyond differences in preferences or information, divergent opinions
emanate from fundamentally different interpretations of reality that take
the form of $stories$. Consequently, a policy gains in popularity if it can
be sustained by an effective narrative; and politicians and public-opinion
makers spend considerable energy on trying to shape the popular narratives
that surround policy debates.

There are countless expressions of this idea in popular and academic
discourse. For instance, a recent New Yorker profile of a former aide of
President Obama begins with the words \textquotedblleft Barack Obama was a
writer before he became a politician, and he saw his Presidency as a
struggle over narrative\textquotedblright .\footnote{%
See
https://www.newyorker.com/magazine/2018/06/18/witnessing-the-obama-presidency-from-start-to-finish.%
} Likewise, two public policy professors write in an LSE blog that
\textquotedblleft there can be little doubt then that people think
narratives are important and that crafting, manipulating, or influencing
them likely shapes public policy\textquotedblright . They add that
narratives simplify complex policy issues \textquotedblleft by telling a
story that includes assertions about what causes what, who the victims are,
who is causing the harm, and what should be done\textquotedblright .%
\footnote{%
See
http://blogs.lse.ac.uk/impactofsocialsciences/2018/07/18/mastering-the-art-of-the-narrative-using-stories-to-shape-public-policy/.%
}

In this paper we offer a formalization of the idea that battles over public
opinion involve competing narratives. Of course, the term \textquotedblleft
narrative\textquotedblright\ is vague and any formalization inevitably
leaves many of its aspects outside the scope of investigation. Our model is
based on the idea that in the context of public-policy debates, narratives
can be regarded as \textit{causal models} that map actions to consequences.
Following the literature on probabilistic graphical models in Statistics,
Artificial Intelligence and Psychology (Cowell et al. (1999), Sloman (2005),
Pearl (2009)), we represent such causal models by directed acyclic graphs
(DAGs).

In our model, what defines a narrative is the variables it incorporates and
the way these are arranged in the causal mapping from actions to
consequences. For instance, consider a debate over US trade policy and its
possible implications for employment in the local manufacturing sector.
Suppose that the public has homogenous preferences over actions and
consequences; disagreements only arise from different beliefs. The DAG 
\begin{equation}
\text{trade policy \ }\rightarrow \text{ \ imports from China \ }\rightarrow 
\text{ \ employment}  \label{DAG_lever_trade}
\end{equation}%
represents a narrative that weaves a third variable (imports from China)
into a causal story that regulates the action-consequence mapping which is
the subject of the policy debate.

The nodes in the DAG represent variables (not the values they can take), and
the links represent perceived direct causal effects (but not the sign or
magnitude of these effects). The variables are coarse-grained, such that the
narrative does not describe an individual historical episode; instead, it
can be used to interpret a wealth of historical episodes. It alerts the
public's attention to long-run correlations between adjacent variables along
the causal chain and invites a causal interpretation of these correlations.

We refer to the narrative represented by (\ref{DAG_lever_trade}) as a
\textquotedblleft \textit{lever narrative}\textquotedblright\ because it
regards imports from China as a \textquotedblleft lever\textquotedblright\ -
i.e., as an endogenous variable that is influenced by policy and in turn
influences the target variable. Intuitively, this narrative supports a
protectionist policy: imports from China are negatively correlated with both
protectionism and employment in the local manufacturing sector, and it is
natural to interpret these correlations in terms of the causal chain (\ref%
{DAG_lever_trade}). But while the support is intuitive, it is illusory if
the narrative is false - e.g. if the actual correlation between imports from
China and employment is due to the confounding effect of exogenous
technological change.

The following is another example of a lever narrative in the context of a
foreign policy debate. The policy question is whether to impose economic
sanctions on a rival country with a hostile regime. The public considers
destabilizing the regime a desirable outcome. A lever narrative that
intuitively gives support to a hawkish policy is%
\[
\text{sanction policy \ }\rightarrow \text{ \ economic situation in rival
country \ }\rightarrow \text{ \ regime\ stability} 
\]%
The following is a lever narrative that involves a different
\textquotedblleft lever\textquotedblright :%
\[
\text{sanction policy \ }\rightarrow \text{ \ nationalism in rival country \ 
}\rightarrow \text{ \ regime\ stability} 
\]%
This narrative intuitively supports a $dovish$ policy because nationalistic
sentiments in the rival country are positively correlated with the stability
of its regime and potentially ameliorated by a soft stance on sanctions.

Thus, two narratives may have the same \textquotedblleft
lever\textquotedblright\ structure but differ in the selection of variables
that function as \textquotedblleft levers\textquotedblright , and
consequently in the policies they support. Likewise, the same variable can
be assigned different roles in the causal scheme. For instance, the
following is a foreign-policy narrative that treats nationalism as an $%
exogenous$ variable:%
\[
\text{sanction policy \ }\rightarrow \text{ \ regime\ stability \ }%
\leftarrow \text{ \ nationalism in rival country} 
\]%
We refer to a narrative with this structure as a \textquotedblleft
threat/opportunity narrative\textquotedblright , because it regards the
third variable that it weaves into the story as an external variable that
the policy $responds$ to rather than influences it. In the context of our
foreign-policy example, this narrative intuitively favors a hawkish policy
because it regards the prospect of waning nationalism in the rival country
as an opportunity for toppling its regime, which tough sanction policy can
exploit.

Thus, foreign-policy narratives can differ in the variables they weave into
the story or in the role that these variables play in the causal mapping
from actions to consequences. This is akin to a dramatist's decision about
which events to include as ingredients in a story and how to construct a
plot around them. Different narratives can generate different beliefs
regarding the mapping from actions to consequences - and therefore lend
support to different policies - because they alert the audience's attention
to correlations between different sets of variables and manipulate its
causal interpretation of these correlations. A public-opinion maker who
wishes to promote a particular policy will therefore devise a narrative that
\textquotedblleft sells\textquotedblright\ it most effectively.

Our objective is to define a notion of equilibrium in public-policy debates,
in which narrative-policy pairs vie for dominance in public opinion. When
the public adopts a narrative, we assume - following Spiegler (2016) - that
it constructs a belief over the narrative's variables, by factorizing their
objective joint distribution according to the so-called \textquotedblleft
Bayesian-Network factorization formula\textquotedblright , and it relies on
this belief to evaluate policies. This factorization captures the notion of
fitting the causal model to objective data. A wrong causal model can induce
a distorted belief regarding the mapping from actions to consequences.

To summarize the first ingredient of our model, a narrative is an
arrangement of selected variables in a causal model (formalized as a DAG),
combined with a rule for generating beliefs from such a causal model. But
what happens when the public confronts competing narratives? Here we invoke
the second ingredient of our model, which is the idea that the public
selects between narrative-policy pairs \textquotedblleft
hedonically\textquotedblright\ - i.e., according to the indirect
anticipatory utility that each one of them generates.

The idea that people adopt distorted beliefs to enhance their anticipatory
utility has several precedents in the literature (Akerlof and Dickens 1982),
Brunnermeier and Parker (2005), Spiegler (2008)). Recently, Montiel Olea et
al. (2018) studied the notion of \textquotedblleft competing
models\textquotedblright\ in a very different context of linear regression
models that differ in the set of variables they admit, and assumed that
prevailing models maximize the indirect expected utility they induce when
estimated against a random sample. In the context of public-policy debates,
we find it particularly natural to assume that the public will be drawn to 
\textit{hopeful} narrative-policy pairs. Precisely because individuals have
little influence over public policy, they incur negligible decision costs
when indulging in hopeful fantasies. It is therefore realistic to assume
that anticipatory feelings are a powerful driving force behind political
positions.

Based on these two ingredients, we define equilibrium as a steady-state
distribution over narrative-policy pairs, such that every element in the
support maximizes a representative agent's anticipatory utility. Why we do
refer to this concept as \textquotedblleft equilibrium\textquotedblright\
instead of plain maximization? The reason is that the action frequencies
that are induced by a given distribution over narrative-policy pairs can
affect belief (and hence the anticipatory utility) that each narrative
generates. This feedback effect is fundamental to the idea of beliefs that
are generated by fitting a wrong causal model to objective long-run data
(see Spiegler (2016)), and it is what creates the need for an equilibrium
approach to the notion of competing narratives.

We employ our equilibrium concept to explore several questions: Which
narratives are attached to various policies - that is, what is their causal
structure and what kind of variables do they involve? Can we account for
divergent popular policies by the notion of competing narratives? Are swings
between conflicting dominant narratives fundamental to battles over public
opinion? The results we present demonstrate the formalism's potential to
shed light on the role of narratives in political debates.\medskip

\noindent \textit{Related literature}

\noindent The idea that people think about empirical regularities in terms
of \textquotedblleft causal stories\textquotedblright\ that can be
represented by DAGs has been embraced by psychologists of causal reasoning
(e.g. Sloman (2005), Sloman and Lagnado (2015)). Spiegler (2016) adopted
this idea as a basis for a model of decision making under causal
misperceptions. In Spiegler (2016), a decision maker forms a subjective
belief by fitting a subjective causal model to objective long-run data. This
continues to be a building block of the model in this paper, which goes
beyond it in two major directions: first, the collection of variables that
can appear in a causal model of a given size is not fixed but selected
endogenously; and second, we assume \textquotedblleft
hedonic\textquotedblright\ selection between competing causal models.

We are aware of at least three papers in economics that draw attention to
the role of narratives in economic contexts. Given that the term
\textquotedblleft narrative\textquotedblright\ has such a loose meaning, it
should come as no surprise that it has received very different
formalizations. Shiller (2017) does not provide an explicit model of what a
narrative is. Instead, he regards certain terms and expressions that appear
in popular discourse as indications of a specific narrative and proposes to
use epidemiological models to study their spread. Benabou et al. (2016)
focus on moral decision making and formalize narratives as messages or
signals that can affect decision makers' beliefs regarding the externality
of their actions. Levy and Razin (2018) use the term \textquotedblleft
narrative\textquotedblright\ to describe information structures in
game-theoretic settings that people postulate to explain observed behavior.

Finally, our paper joins a handful of works in so-called \textquotedblleft
behavioral political economics\textquotedblright\ that study voters' belief
formation according to misspecified subjective models or wrong causal
attribution rules - e.g., Spiegler (2013), Esponda and Pouzo (2017); and see
Schnellenbach and Schubert (2015) for a survey.

\section{The Model}

Let $X=X_{1}\times \cdots \times X_{n}$, where $n>2$ and $X_{i}=\{0,1\}$ for
each $i=1,...,n$. For every $N\subseteq \{1,...,n\}$, denote $X_{N}=\times
_{i\in N}X_{i}$. For any $x\in X$, the components $x_{1}$ and $x_{n}$ - also
denoted $a$ and $y$ - are referred to as an $action$ and a $consequence$.
These components are independently distributed. In particular, actions have
no causal effects on consequences.

Let $Q$ be a finite set of conditional distributions over $x_{2},...,x_{n-1}$
that have full support for every $x_{1},x_{n}$. Given a pair of numbers $%
\alpha ,\mu \in (0,1)$, define $P_{\alpha ,\mu }\subset \Delta (X)$ as the
set of distributions $p$ for which $p(a=1)=\alpha $, $p(y=1\mid a)=\mu $ for
all $a$, and $(p(\cdot \mid x_{1},x_{n}))$ is in $Q$. We regard $\mu $ as a
constant, whereas $\alpha $ represents a historical action frequency that we
endogenize below.

A \textit{directed acyclic graph} (DAG) is a pair $(N,R)$, where $N\subseteq
\{1,...,n\}$ is a set of nodes and $R\subseteq N\times N$ is a set of
directed links. Acyclicity means that the graph contains no directed path
from a node to itself. We use $iRj$ or $i\rightarrow j$ to denote a directed
link from the node $i$ into the node $j$. Abusing notation, let $R(i)=\{j\in
N\mid jRi\}$ be the set of \textquotedblleft parents\textquotedblright\ of
node $i$. We will often suppress $N$ in the notation of a DAG and identify
it with $R$.

Following Pearl (2009), we interpret a DAG as a \textit{causal model}, where
the link $i\rightarrow j$ means that $x_{i}$ is perceived as an immediate
cause of $x_{j}$. Directedness and acyclicity of $R$ are consistent with
basic intuitions regarding causality. The causal model is agnostic about the
sign or magnitude of causal effects.

Let $\mathcal{R}$ be a collection of DAGs $(N,R)$ satisfying two
restrictions: $\{1,n\}\subseteq N$, and there is no directed path from $n$
to $1$ - i.e., the consequence variable is not perceived as a (possibly
indirect) cause of the action. In all the DAGs that appear in the examples
we will examine, $1$ is an \textit{ancestral} node (i.e., $R(1)=\varnothing $%
) and $n$ is the unique \textit{terminal} node (i.e., $n\notin R(i)$ for
every $i\in N$ and there is no other node with this property). However,
these properties are not necessary for our general analysis.\medskip

\noindent \textit{Narratives and their induced beliefs}

\noindent Fix $\alpha ,\mu \in (0,1)$. A \textit{narrative} is a pair $%
s=(p,R)\in P_{\alpha ,\mu }\times \mathcal{R}$. The narrative induces a
subjective belief over $\Delta (X_{N})$, defined as follows:%
\begin{equation}
p_{R}(x_{N})=\prod_{i\in N}p(x_{i}\mid x_{R(i)})  \label{DAG}
\end{equation}%
The full-support assumption ensures that all the terms in this factorization
formula are well-defined.

The conditional distribution of $x_{n}$ given $x_{1}$ induced by $p_{R}$ is
computed in the usual way. It has a simple expression when $1$ is an
ancestral node:%
\begin{equation}
p_{R}(x_{n}\mid x_{1})=\sum_{x_{2},...,x_{n-1}}\left( \prod_{i>1}p(x_{i}\mid
x_{R(i)})\right)  \label{conditional_DAG}
\end{equation}%
For illustration, when the DAG is $R:1\rightarrow 3\rightarrow 4\leftarrow 2$%
, the narrative $(p,R)$ induces%
\[
p_{R}(x_{1},x_{2},x_{3},x_{4})=p(x_{1})p(x_{2})p(x_{3}\mid x_{1})p(x_{4}\mid
x_{2},x_{3}) 
\]%
and%
\[
p_{R}(x_{4}\mid x_{1})=\sum_{x_{2},x_{3}}p(x_{2})p(x_{3}\mid
x_{1})p(x_{4}\mid x_{2},x_{3}) 
\]

The interpretation of this belief formation process is as follows. In a
narrative $(p,R)$, the conditional distribution $p(x_{2},...,x_{n-1}\mid
x_{1},x_{n})$ represents a selection of $n-2$ observable \textit{variables }%
that are incorporated into the story. In other words, every conditional
distribution in $Q$ is implemented by some collection of $n-2$ actual
variables. The component $R$ determines how these variables (some or all of
them) are woven into a causal structure. This is akin to a novelist who
conjures up a collection of events, and then organizes their unfolding
according to a plot. The narrative generates a subjective belief regarding
the mapping from actions to consequences, by alerting the audience's
attention to particular correlations - those that the causal model deems
relevant - and combining them according to the causal model. The
correlations themselves are accurate - i.e., each of the terms in the
factorization formula (\ref{DAG}) is extracted from an objective
distribution (over $a,y$ and the selected additional variables). However,
the way they are combined may lead to distorted belief, such that $%
p_{R}(y=1\mid a)\neq \mu $ for some $a$.\medskip

\noindent \textit{Policies and anticipatory utility}

\noindent Let $D=[\varepsilon ,1-\varepsilon ]$, where $\varepsilon >0$ is
arbitrarily small. A \textit{policy} $d\in D$ is a proposed mixture over
actions, where $d$ is the proposed frequency of playing the action $a=1$.

Given a historical action frequency $\alpha $, a narrative $s=(p,R)$ and a
policy $d$ induce the following \textit{gross anticipatory utility}:%
\begin{equation}
V(s,d\mid \alpha )=d\cdot p_{R}(y=1\mid a=1)+(1-d)\cdot p_{R}(y=1\mid a=0)
\label{gross_anticipatory}
\end{equation}%
Note that $V$ is defined for a given $\alpha $ because the set of feasible
narratives varies with $\alpha $, but also (as we will later see) because
the subjective distribution $p_{R}(y\mid a)$ is $not$ invariant to $\alpha $.

A representative agent has a utility function $u(y,d)=y-C(d-d^{\ast })$,
where $d^{\ast }\in D$ is the agent's ideal policy, and $C$ is a symmetric,
convex cost function that satisfies $C(0)=C^{\prime }(0)=0$. The function $C$
represents the agent's intrinsic disutility he experiences when deviating
from his ideal policy. Note that if the agent had rational expectations, he
would realize that $y$ is independent of $a$ and find no reason to deviate
from $d^{\ast }$. Given $\alpha $, The agent's net anticipatory utility from
the narrative-policy pair $(s,d)$ is%
\begin{equation}
U(s,d\mid \alpha )=V(s,d\mid \alpha )-C(d-d^{\ast })
\label{net_anticipatory}
\end{equation}

One may wonder why there is a need to define policy as a continuous
variable, rather than identifying it with the binary action. The reason, as
usual in these cases, is that we want our model to generate a fine mapping
from the subjective belief $p_{R}(y\mid a)$ to policies. In addition,
certain interesting effects in our model would disappear or become obscured
under a binary-policy specification.\medskip

\noindent \textit{Equilibrium}

\noindent The model's primitives are the exogenous probability of a good
outcome $\mu $, the set of conditional distributions $Q$, the set of
feasible DAGs $\mathcal{R}$ and the cost function $C$. The objects $%
P_{\alpha ,\mu }$, $p_{R}$ and $U$ are derived from these primitives. We are
now ready to define our notion of equilibrium.\medskip

\begin{definition}
An action frequency $\alpha \in \lbrack 0,1]$ and a probability distribution 
$\sigma $ over narrative-policy pairs constitute an equilibrium if two
conditions hold:%
\[
Supp(\sigma )\subseteq \arg \max_{(s,d)\in P_{\alpha ,\mu }\times \mathcal{R}%
\times D}U(s,d\mid \alpha ) 
\]%
and%
\[
\alpha =\sum_{(s,d)}\sigma (s,d)\cdot d 
\]%
\bigskip
\end{definition}

This concept captures a steady-state in the battle over public opinion. The
first condition requires that prevailing narrative-policy pairs are those
that maximize the representative agent's net anticipatory utility, given the
historical action frequency. Thus, public opinion's criterion for selecting
between competing narrative-policy pairs is net anticipatory utility - in
other words, it chooses the narrative it prefers to believe in. This
captures the idea that voters do not adjudicate between narratives using
\textquotedblleft scientific\textquotedblright\ methods; rather, they are
attracted to narratives with a hopeful message. The second condition
requires the historical action frequency to be consistent with the marginal
steady-state distribution over policies. The lower and upper limits on $d$
are thus introduced in order to ensure that $\alpha $ is interior.

The distribution $\alpha $ can be interpreted as a cross-section measurement
of the relative popularity of various policies among the public. However, we
favor an \textquotedblleft ergodic\textquotedblright\ interpretation,
according to which $\alpha $ describes a historical action frequency.
Different policies are ascendant at various points in time. A particular
policy rises to dominance when the narrative that accompanies it appeals to
the public in the sense that the narrative-policy pair maximizes the
public's anticipatory payoff. Over time, as the historical action frequency
changes, so does the anticipatory payoff induced by various narrative-policy
pairs, and therefore a different narrative-policy pair may become dominant.
The distribution $\alpha $ is the average action frequency that results from
the periodic swings between dominant narrative-policy pairs.

The following preliminary result establishes equilibrium existence.\medskip

\begin{proposition}
\label{existence}An equilibrium exists.\medskip
\end{proposition}

Our next basic observation provides a simple rational-expectations
benchmark. If $R$ is a fully connected DAG, or if it contains no directed
path from the ancestral node $1$ to node $n$, then $p_{R}(y\mid a)=\mu $ for
all $a$ - i.e. the agent's belief regarding the mapping from actions to
consequences coincides with rational expectations. In this case, $%
V((p,R),d\mid \alpha )=\mu $ for \textit{every} $p,d$, such that deviating
from the ideal policy $d^{\ast }$ does not produce any kick to anticipatory
utility. If $\mathcal{R}$ only consists of such DAGs, then in any
equilibrium $(\alpha ,\sigma )$, the marginal of $\sigma $ over $d$ (and
therefore $\alpha $) assigns probability one to $d^{\ast }$. In the next
section, we will begin to see departures from this crisp benchmark when
other DAGs are admitted.

\section{An Example: Foreign-Policy Narratives}

Let $n=3$, $\mu =d^{\ast }=\frac{1}{2}$, $C(\Delta )=k\Delta ^{2}$, where $k>%
\frac{\sqrt{2}}{4}$. Take $\varepsilon $ (in the definition of $D$) to be
vanishingly small. Suppose that $Q$ consists of a $single$ conditional
distribution:%
\begin{equation}
p(x_{2}=1\mid a,y)\approx a(1-y)  \label{foreignpolicy}
\end{equation}%
The approximate equality is due to an arbitrarily small perturbation of the
exact specification $x_{2}=a(1-y)$, to ensure that $p$ has full support. The
set $\mathcal{R}$ consists of all DAGs with two or three nodes in which $a$
is represented by a ancestral node.

Interpret the three variables as follows. The action $a$ represents foreign
policy toward a rival country with a hostile regime, where $a=1$ ($0$)
denotes hawkish (dovish) policy. The consequence $y$ represents the
stability of the regime, where $y=1$ ($0$) indicates regime change (regime
stability). Finally, the variable $x_{2}$ represents the strength of
nationalistic attitudes among the rival country's population, where $x_{2}=1$
($0$) indicates that these attitudes are strong (weak).

The joint distribution $p$ satisfies the following properties. First,
foreign policy has no causal effect on the stability of the rival country's
regime. Second, hawkish (dovish) policy tends to strengthen (weaken)
nationalism in the rival country. Finally, nationalism and regime stability
are positively correlated. In particular, regime change can only happen when
nationalistic attitudes are weak. Yet, this correlation is $not$ causal;
rather, it is due to confounding by exogenous variables that are excluded
from the causal models our narrators employ.

Since $Q$ is a singleton in this example, narrators have no freedom in their
choice of $p$. Consequently, a narrative can be identified with the DAG it
employs.\medskip

\begin{claim}
There exists a unique equilibrium $(\alpha ,\sigma )$, where $\alpha \approx
2-\sqrt{2}$ and $Supp(\sigma )$ consists of two narrative-policy pairs: $(i)$
a lever narrative $R^{l}:a\rightarrow x_{2}\rightarrow y$ coupled with a
dovish policy $d^{o}\approx \frac{1}{2}-\frac{1}{8}\frac{\sqrt{2}}{k}$; $%
(ii) $ an opportunity narrative $R^{o}:a\rightarrow y\leftarrow x_{2}$,
coupled with a hawkish policy $d^{l}\approx \frac{1}{2}+\frac{1}{8}\frac{%
\sqrt{2}}{k} $.
\end{claim}

\begin{proof}
For the sake of the calculations in this proof, we treat the
approximate-equality definition of $p$ as if the equality were precise. We
will also suppose that the equilibrium policies are interior and given by
first-order conditions. We will later verify that the equilibrium is unique.

Consider the opportunity DAG $R^{o}$. By (\ref{conditional_DAG}), we have%
\[
p_{R^{o}}(y\mid a)=\sum_{x_{2}=0,1}p(x_{2})p(y\mid a,x_{2}) 
\]%
We can calculate these terms under the specification (\ref{foreignpolicy})
and the assumption that $\mu =\frac{1}{2}$, and obtain%
\begin{eqnarray*}
p_{R^{o}}(y &=&1\mid a=0)=\frac{2-\alpha }{4} \\
p_{R^{o}}(y &=&1\mid a=1)=\frac{2-\alpha }{2}
\end{eqnarray*}%
such that%
\begin{equation}
U(R^{o},d\mid \alpha )=d\cdot \frac{2-\alpha }{2}+(1-d)\cdot \frac{2-\alpha 
}{4}-k(d-\frac{1}{2})^{2}  \label{exampleexpression1}
\end{equation}%
Therefore,%
\begin{equation}
\frac{\partial U(R^{o},d\mid \alpha )}{\partial d}=\frac{2-\alpha }{4}-2k(d-%
\frac{1}{2})  \label{exampleexpression2}
\end{equation}%
Because this derivative is strictly positive at $d\leq \frac{1}{2}$ and
strictly decreasing in $d>\frac{1}{2}$, there is a unique policy $d^{o}>%
\frac{1}{2}$ that maximizes $U(R^{o},d\mid \alpha )$.

Now consider the lever DAG $R^{l}$. By (\ref{conditional_DAG}), we have%
\[
p_{R^{l}}(y\mid a)=\sum_{x_{2}=0,1}p(x_{2}\mid a)p(y\mid x_{2}) 
\]%
We can calculate these terms under the specification (\ref{foreignpolicy})
and the assumption that $\mu =\frac{1}{2}$, and obtain%
\begin{eqnarray*}
p_{R^{l}}(y &=&1\mid a=0)=\frac{1}{2-\alpha } \\
p_{R^{l}}(y &=&1\mid a=1)=\frac{1}{2(2-\alpha )}
\end{eqnarray*}%
such that%
\begin{equation}
U(R^{l},d\mid \alpha )=d\cdot \frac{1}{2(2-\alpha )}+(1-d)\cdot \frac{1}{%
2-\alpha }-k(d-\frac{1}{2})^{2}  \label{examplexpression3}
\end{equation}%
Therefore,%
\begin{equation}
\frac{\partial U(R^{l},d\mid \alpha )}{\partial d}=-\frac{1}{2(2-\alpha )}%
-2k(d-\frac{1}{2})  \label{examplexpression4}
\end{equation}%
Because this derivative is strictly negative at $d\geq \frac{1}{2}$ and
strictly decreasing in $d>\frac{1}{2}$, there is a unique policy $d^{l}<%
\frac{1}{2}$ that maximizes $U(R^{l},d\mid \alpha )$. It follows that $%
Supp(\sigma )$ must be some weak subset of $\{(R^{o},d^{o}),(R^{l},d^{l})\}$.

Let us first suppose that $Supp(\sigma )$ coincides with this set and that $%
d^{o}$ and $d^{l}$ are given by first-order conditions. Then,%
\begin{eqnarray}
U(R^{o},d^{o} &\mid &\alpha )=U(R^{l},d^{l}\mid \alpha )  \label{U(RO)=U(RL)}
\\
\frac{\partial U(R^{o},d\mid \alpha )\mid _{d=d^{o}}}{\partial d} &=&\frac{%
\partial U(R^{l},d\mid \alpha )\mid _{d=d^{l}}}{\partial d}=0  \label{FOC}
\end{eqnarray}%
By plugging (\ref{exampleexpression1})-(\ref{examplexpression4}) into the
above equations, we can verify that they are satisfied at the values for $%
(d^{o},d^{l},\alpha )$ that are given in the statement of the claim. The
assumption on $k$ ensures that the solution is well-defined. The exact
weights that $\sigma $ assigns to the two points in the support can be
extracted from the condition $\alpha =\sum_{(s,d)}\sigma (s,d)\cdot d$.

To verify uniqueness, consider first equilibria in which $Supp(\sigma )$ has
two elements. Note that $U(R^{o},d^{o}\mid \alpha )$ monotonically \textit{%
decreases} with $\alpha $, while $U(R^{l},d^{l}\mid \alpha )$ monotonically 
\textit{increases} with $\alpha $. This means that for a given $%
(d^{o},d^{l}),$ there is a unique $\alpha $ that solves equation (\ref%
{U(RO)=U(RL)}). Given $\alpha ,$ equations (\ref{U(RO)=U(RL)})-(\ref{FOC})
are linear in $(d^{o},d^{l})$ and hence, have a unique solution. It follows
that there is a unique triplet $(d^{o},d^{l},\alpha )$ that solves (\ref%
{U(RO)=U(RL)})-(\ref{FOC}). Now suppose that $Supp(\sigma )$ consists of a
single point $(R^{l},d)$ ($(R^{o},d)$) only. Then, $\alpha =d$. In this
case, a simple calculation establishes that the narrative-policy pair $%
(R^{o},1-d)$ ($(R^{l},1-d)$) delivers a higher net anticipatory utility, a
contradiction.\medskip
\end{proof}

This example has a number of noteworthy features.\medskip

\noindent \textit{Coupling of narratives and policies}

\noindent Although there is a single available variable (other than the
action and the consequence) that narrators can incorporate into their
stories, its location in the narrative's causal scheme depends on the
direction of the policy the narrative is meant to sustain. Thus, in order to
sustain a hawkish policy $d>d^{\ast }$, the narrative must treat the
variable $x_{2}$ as an exogenous opportunity. In contrast, to sustain a
dovish policy $d<d^{\ast }$, the narrative must treat the variable $x_{2}$
as a lever.

The reason that the lever narrative promotes dovish policies is that
according to $p$, $a$ and $x_{2}$ are positively correlated, whereas $x_{2}$
and $y$ are negatively correlated. The lever narrative puts these
correlations together as if they reflected a causal chain $a\rightarrow
x_{2}\rightarrow y$. As a result, $p_{R^{l}}$ predicts a negative indirect
causal effect of $a$ on $y$.

The intuition for why the opportunity narrative promotes hawkish policies is
quite different. According to $p$, $\Pr (a=1,x_{2}=0)\approx \alpha \mu $ -
i.e., the combination of $a=1$ and $x_{2}=0$ is an infrequent event. Yet the
rarity is unaccounted for by $p_{R^{o}}$, which sums over $x_{2}$ without
conditioning on $a$ (and observe that $\Pr (x_{2}=0)=\alpha \mu +1-\alpha
>\Pr (a=1,x_{2}=0)$). At the same time, the probability of $y=1$ conditional
on the combination $a=1,x_{2}=0$ is approximately one: if we observe both
hawkish policy and weak nationalism, it is almost surely because the regime
is unstable. The coupling of these two effects leads to an exaggerated
belief in the probability of $y=1$ conditional on $a=1$.\medskip

\noindent \textit{Equilibrium polarization}

\noindent The marginal equilibrium distribution over policies assigns weight
to one policy on each side of the agent's ideal point. The fundamental force
behind this polarization effect is a \textquotedblleft diminishing
returns\textquotedblright\ property of the two narratives: their ability to
deceive the agent about the effect of $a$ on $y$ decreases with the
historical frequency of the action they support. Thus, when we perturb $%
\alpha $ above the equilibrium level, this makes room for the growing
popularity of a lever narrative that sustains a dovish policy. Conversely,
perturbing $\alpha $ below the equilibrium level increases the popularity of
an opportunity narrative that promotes a hawkish policy.

This effect can be interpreted in terms of cross-sectional political
polarization: At any moment in time, there are two narrative-policy pairs
that dominate public opinion. Alternatively, it can be given an
\textquotedblleft ergodic\textquotedblright\ interpretation: Different
narrative-policy pairs rise to dominance at different points in time, and
the distribution $\sigma $ captures the long-run frequency with which each
of them is dominant.\medskip

\noindent \textit{Mutual narrative refutation}

\noindent In our model, the representative agent does not reason
\textquotedblleft scientifically\textquotedblright\ about the causal models
conveyed by conflicting narratives. Rather than actively seeking data about $%
p(y\mid a)$ in order to test the contending narratives, he allows the
\textquotedblleft narrators\textquotedblright\ to determine the data he pays
attention to. Thus, the lever narrative calls his attention to the
conditional probabilities $p(x_{2}\mid a)$ and $p(y\mid x_{2})$, whereas the
opportunity narrative calls his attention to the marginal probability $%
p(x_{2})$ and the conditional probability $p(y\mid a,x_{2})$. When
evaluating a given narrative $(p,R)$, the agent only considers the data that
the narrative calls attention to and uses it to evaluate the narrative's
anticipatory value, via the factorization formula $p_{R}$.

If our agent were somewhat less passive in his approach to data, he could
notice that the data that one narrative employs actually refutes the other
narrative. Thus, the data $p(y\mid a,x_{2})$ referred to by the opportunity
narrative demonstrates that unlike what the lever narrative assumes, $y$ and 
$a$ are $not$ independent conditional on $x_{2}$. Conversely, the data $%
p(x_{2}\mid a)$ demonstrates that unlike what the opportunity narrative
assumes, $x_{2}$ and $a$ are $not$ independent. But how would the agent
respond to this observation? A critical reaction would be to distrust all
narratives and develop a more \textquotedblleft
scientific\textquotedblright\ belief-formation method. However, an equally
natural reaction would be to conclude that \textquotedblleft all narratives
are wrong\textquotedblright\ and stick to the one that makes the agent feel
more hopeful about the future - especially in the political context, where
the agent's personal stakes are negligible.

Finally, note that this scenario would not arise in a modified version of
our example, in which there are $two$ distinct variables with the same
conditional distribution. In this case, the two conflicting narratives could
invoke different variables, such that the above mutual refutation would be
infeasible.\medskip

\noindent \textit{Hawkish bias and distortion of the status quo}

\noindent For a given absolute policy distance from the ideal point $d^{\ast
}=\frac{1}{2}$, the opportunity narrative leads to a higher anticipatory
utility than the lever narrative. As a result, the average equilibrium
policy lands on the hawkish side (even though $d^{o}$ and $d^{l}$ are
equally far from the ideal point) - i.e., $\alpha >\frac{1}{2}$.

The fundamental reason behind this effect is that given $p$, the lever
narrative has the property that $V((p,R^{l}),\alpha \mid \alpha )=\mu $,
whereas the opportunity narrative satisfies $V((p,R^{l}),\alpha \mid \alpha
)>\mu $. In other words, while the lever narrative exaggerates the
probability of $y=1$ under a $counterfactual$ dovish movement away from the
steady-state policy, it does $not$ distort the consequences of a policy that
adheres to the status quo. In contrast, the opportunity narrative also
distorts the status-quo.

This ability to spin tales not just about counterfactual events but also
about the status quo gives the opportunity narrative an advantage over the
lever narrative. A plausible criterion for refining our notion of
equilibrium is to rule out such distortions of the status quo because the
public is less likely to fall for a narrative that misrepresents the status
quo. Our analysis in the next section will involve such a restriction. In
the current example, it rules out the opportunity narrative (in fact, this
is generically the case). The following result summarizes the effect of this
change on the equilibrium analysis.\medskip

\begin{claim}
Suppose that $\mathcal{R}$ includes all the DAGs in the original
specification except $a\rightarrow y\leftarrow x_{2}$. Then, there exists an
essentially unique equilibrium $(\alpha ,\sigma )$, where $\alpha \approx 
\frac{5}{4}-\frac{1}{4}\sqrt{9+\frac{2}{k}}$, and $Supp(\sigma )$ consists
of the following narrative-policy pairs: $(i)$ a lever narrative $%
R^{l}:a\rightarrow x_{2}\rightarrow y$ coupled with a dovish policy $%
d^{l}\approx 2-\frac{1}{2}\sqrt{9+\frac{2}{k}}$; $(ii)$ any distribution
over the remaining DAGs in $\mathcal{R}$ coupled with the policy $d^{\ast }$%
.\medskip
\end{claim}

The proof follows the same outline as in the previous claim, except that the
policy $d^{\ast }$ coupled with any DAG that induces rational expectations
(e.g. $a\rightarrow y$) replaces $(R^{o},d^{o})$. Thus, when the opportunity
narrative is ruled out, the equilibrium exhibits a dovish bias, mixing
between the rational-expectations policy $d^{\ast }$ and a dovish policy
that is sustained by the lever narrative.

\section{Analysis}

Toward the end of the previous section, we pointed out that while narratives
distort the effect of $a$ on $y$, a plausible restriction is that this
distortion only involves $counterfactual$ deviations from the steady-state
policy. It is one thing to stoke illusions about the consequences of
counterfactual policies, and quite another to present a wrong picture about
the consequences of actual policies, because the latter can be checked
against the long-run observation of $p(y)$. Hence, it seems sensible to
restrict attention to narratives that do not distort beliefs about the
effectiveness of the status-quo policy. In this section, we implement this
desideratum by restricting the set of feasible DAGs $\mathcal{R}$.\medskip

\begin{definition}[Perfect DAGs]
A DAG $(N,R)$ is perfect if whenever $iRk$ and $jRk$ for some $i,j,k\in N$,
it is the case that $iRj$ or $jRi$.\medskip
\end{definition}

Thus, in a causal model that is represented by a perfect DAG, if two
variables are perceived as direct causes of a third variable, then there
must be a perceived direct causal link between them. E.g., $1\rightarrow
2\rightarrow 3$ is perfect, and so is the more elaborate DAG:%
\begin{equation}
\begin{array}{ccccccc}
1 & \rightarrow & 2 & \rightarrow & 4 & \rightarrow & 6 \\ 
& \searrow & \downarrow & \nearrow & \downarrow & \nearrow &  \\ 
&  & 3 & \rightarrow & 5 &  & 
\end{array}
\label{perfectDAGexamples}
\end{equation}%
In contrast, the DAG $1\rightarrow 3\leftarrow 2$ is imperfect because $1R3$
and $2R3$, yet there is no direct link between $1$ and $2$.

Perfection is a familiar property in the Bayesian Networks literature. In
our context, the crucial properties of perfect DAGs are the
following:\medskip

\noindent \textit{Correct marginals}. Let $(N,R)$ be a perfect DAG. Then, $%
p_{R}(x_{i})=p(x_{i})$ for every $i\in N$. That is, the subjective
distribution induced by the DAG does not distort the objective marginal
distribution over individual variables.\medskip

\noindent \textit{No status-quo distortion} (NSQD). Let $(N,R)$ be a perfect
DAG. Then, $V((p,R),\alpha \mid \alpha )=\mu $ for every objective
distribution $p$. That is, the DAG never distorts the consequences of
following a policy that coincides with the historical action
frequencies.\medskip

Indeed, Spiegler (2017,2018) shows that the class of perfect DAGs is the
largest that satisfies these properties for all objective distributions.
This observation can be extended: For a $generic$ $p$, imperfect DAGs will
violate both properties. Thus, the significance of the restriction to
perfect DAGs is that it is necessary for the NSQD property, given a generic
set $Q$.

\subsection{Linear Narratives}

In this sub-section we investigate the structure of narratives.
Specifically, we focus on the notion of linear DAGs.\medskip

\begin{definition}
A DAG $(N,R)$ is linear if $1$ is the unique ancestral node, $n$ is the
unique terminal node, and $R(i)$ is a singleton for every non-ancestral
node.\medskip
\end{definition}

Clearly, linear DAGs are a subclass of perfect DAGs, because by definition,
no node in a linear DAG has more than one parent. Linear DAGs capture the
simplest form of narrative. They consist of a single causal chain and
correspond to the notion of stories as \textquotedblleft one damned thing
after another\textquotedblright . In addition, they are simple in the sense
that they only call attention to correlations between $pairs$ of variables
(this property characterizes any causal tree - indeed, linear DAGs are
degenerate trees with a single terminal node).

The intuitive appeal of linear DAGs raises the question of whether there is
any loss of generality in restricting attention to them. Formally, we pose
the following question. Consider a narrative $(p,R)$ in which $R$ is a
perfect DAG. Is there an alternative narrative $(p^{\prime },R^{\prime })$
in which $R^{\prime }$ is linear (and not larger than $R$, in the sense that
it has weakly fewer nodes), such that $p_{R^{\prime }}^{\prime }(y\mid
a)=p_{R}(y\mid a)$?

Looking at the illustrative perfect DAGs at the beginning of this section,
one might get the impression that the answer is obvious. For instance, in
the DAG given by (\ref{perfectDAGexamples}), we could collapse the subsets $%
\{2,3\}$ and $\{4,5\}$ into a pair of "mega-nodes" $x_{2}^{\prime
}=(x_{2},x_{3})$ and $x_{4}^{\prime }=(x_{4},x_{5})$, such that the six-node
perfect DAG, denoted $R$, would be reduced to a four-node linear DAG $%
R^{\prime }:1\rightarrow 2^{\prime }\rightarrow 4^{\prime }\rightarrow 6$.
However, note that for a given $p$, the original DAG $R$ induces%
\[
p_{R}(x_{1},...,x_{6})=p(x_{1},x_{2},x_{3})p(x_{4}\mid
x_{2},x_{3})p(x_{5}\mid x_{3},x_{4})p(x_{6}\mid x_{4},x_{5}) 
\]%
whereas the reduced DAG leads to a factorization that can be written as%
\[
p_{R^{\prime }}(x_{1},...,x_{6})=p(x_{1},x_{2},x_{3})p(x_{4}\mid
x_{2},x_{3})p(x_{5}\mid x_{2},x_{3},x_{4})p(x_{6}\mid x_{4},x_{5}) 
\]%
The third terms in these two expressions are different. Therefore, for
arbitrary $p$, we will have $p_{R^{\prime }}\neq p_{R}$ and it is not
immediately obvious that we could come up with a different $p^{\prime }$
such that $p_{R^{\prime }}^{\prime }(x_{6}\mid x_{1})=p_{R}(x_{6}\mid x_{1})$%
.\medskip

\begin{proposition}
\label{prop_linearDAG}For every narrative $(p,R)$ in which $R$ is perfect,
there exists another narrative $(p^{\prime },R^{\prime })$ in which $%
R^{\prime }$ is linear and has weakly fewer nodes than $R$, such that $%
p_{R^{\prime }}^{\prime }(y\mid a)\equiv p_{R}(y\mid a)$.\medskip
\end{proposition}

Thus, for every narrative $(p,R)$ that employs a perfect DAG we can find a
(potentially different) narrative $(p^{\prime },R^{\prime })$ in which $%
R^{\prime }$ is a linear DAG with weakly fewer nodes than $R$, such that the
two narratives generate the same conditional beliefs. The intermediate nodes
in $R^{\prime }$ represent variables that are derived from the original
variables via a non-trivial sequence of transformations, which employs the
basic tool of \textquotedblleft junction trees\textquotedblright\ in the
Bayesian Networks literature. Therefore, $p^{\prime }$ is typically
different from $p$. In particular, this means that $p^{\prime }$ may lie
outside the set $Q$ to which $p$ belongs. That is, our result does $not$
mean that the restriction to linear DAGs is without loss of generality for
an \textit{arbitrary} set $Q$. However, if $Q$ is sufficiently rich, linear
narratives can approximate non-linear narratives that involve perfect DAGs.

\subsection{Polarization}

As shown at the end of Section 2, under rational expectations (or when $%
\mathcal{R}$ only consists of DAGs that induce $p_{R}(y=1\mid a)=\mu $ for
all $a$), any equilibrium assigns probability one to the ideal policy $%
d^{\ast }$. This provides a stark benchmark for the result in this
sub-section.\medskip

\begin{definition}
Fix $\mu $. A pair $(Q,\mathcal{R)}$ is $rich$ if it satisfies the following
two conditions: (i) for every $\alpha \in (0,1)$ there exists a feasible
narrative $(p,R)$, $p\in P_{\alpha ,\mu }$, $R\in \mathcal{R},$ such that $%
p_{R}(y=1\mid a)$ is non-constant in $a$, and (ii) for every $q\in Q$ there
exists $q^{\prime }\in Q$ such that $q^{\prime }(\cdot \mid a,y)\equiv
q(\cdot \mid 1-a,y)$.\medskip
\end{definition}

Richness means that the set of feasible narratives always enables belief
distortions that favor either action. To see why it is not a vacuous
property, recall that the lever narrative in Section 3 satisfies $%
p_{R}(y=1\mid a=0)>p_{R}(y=1\mid a=1)$. Because $Q$ is a singleton in that
example, it fails condition $(ii)$ in the definition of richness. Now add to 
$Q$ a mirror image of the conditional distribution given by (\ref%
{foreignpolicy}), such that $x_{2}=(1-a)(1-y)$ with arbitrarily high
probability. Then, as long as $\mathcal{R}$ includes $a\rightarrow
x_{2}\rightarrow y$, the pair $(Q,\mathcal{R})$ is rich.\medskip

\begin{proposition}
Let $\mathcal{R}$ be a collection of perfect DAGs, such that $(Q,\mathcal{R)}
$ is rich. Then, in any equilibrium $(\alpha ,\sigma )$, $\sigma $ assigns
positive probability to exactly two policies, $d_{r}>d^{\ast }$ and $%
d_{l}<d^{\ast }$.
\end{proposition}

\begin{proof}
Fix an equilibrium $(\alpha ,\sigma )$. First, we establish that the support
of $\sigma $ must include least two distinct policies. Assume the contrary -
i.e., the marginal of $\sigma $ over $d$ is degenerate. Then by definition,
it assigns probability one to the steady-state policy $\alpha $. By the NSQD
property of perfect DAGs, $V(s,\alpha \mid \alpha )=\mu $ for every feasible
narrative $s$.

There are two cases to consider. Suppose $\alpha \neq d^{\ast }$. Then any
narrative $(p,R)$ in the support of $\sigma $ delivers $U((p,R),d^{\ast
}\mid \alpha )=\mu -C(\alpha -d^{\ast })$. However, the narrative policy
pair $((p,R^{\ast }),d^{\ast })$, where $R^{\ast }=a\rightarrow y$ generates
the net payoff $U((p,R^{\ast }),d^{\ast }\mid \alpha )=\mu $, contradicting
the first part of the definition of equilibrium. Suppose next that $\alpha
=d^{\ast }$. Then,%
\[
V((p,R),d^{\ast }\mid \alpha )=d^{\ast }\cdot p_{R}(y=1\mid a=1)+(1-d^{\ast
})\cdot p_{R}(y=1\mid a=0)=\mu 
\]%
By property $(i)$ of richness, there is a feasible narrative $(p^{\prime
},R^{\prime })$ such that without loss of generality, $p_{R^{\prime
}}^{\prime }(y=1\mid a=1)>p_{R^{\prime }}^{\prime }(y=1\mid a=0)$. Therefore,%
\[
V((p^{\prime },R^{\prime }),d^{\prime }\mid \alpha )=d^{\prime }\cdot
p_{R}(y=1\mid a=1)+(1-d^{\prime })\cdot p_{R}(y=1\mid a=0)>\mu 
\]%
whenever $d^{\prime }>d^{\ast }$. Since $C^{\prime }=0$ at $d=d^{\ast }$, it
follows that coupling the narrative $(p^{\prime },R^{\prime })$ with such a
policy $d^{\prime }$ that is slightly larger than $d^{\ast }$ will deliver $%
U((p^{\prime },R^{\prime }),d^{\prime })>\mu $, a contradiction.

Now suppose that the support of $\sigma $ contains at least two distinct
policies. We argue that at least two of these policies, denoted $d_{l}$ and $%
d_{r},$ satisfy $d_{l}<\alpha $ and $d_{r}>\alpha $. Note that every $%
(s,d)\in Supp(\sigma )$ must deliver $U(s,d)\geq \mu $ because the
narrative-policy pair $((p,a\rightarrow y),d^{\ast })$ induces $U=\mu $. Let
us now show that the narrative $(p_{1},R_{1})$ that accompanies the policy $%
d_{r}$ satisfies $p_{R_{1}}(y=1\mid a=1)>p_{R_{1}}(y=1\mid a=0)$, and that
the narrative $(p_{0},R_{0})$ that accompanies $d_{l}$ satisfies $%
p_{R_{0}}(y=1\mid a=1)<p_{R_{0}}(y=1\mid a=0)$.

By the definition of equilibrium, any narrative $(p,R)$ that accompanies\
any $d$ in the support of $\sigma $ maximizes%
\[
U((p,R),d\mid \alpha )=V((p,R),d\mid \alpha )-C(d-d^{\ast }) 
\]%
where%
\[
V((p,R),d\mid \alpha )=d\cdot p_{R}(y=1\mid a=1)+(1-d)\cdot p_{R}(y=1\mid
a=0) 
\]%
Because all feasible narratives involve perfect DAGs, any $(p,R)$ must
satisfy $V((p,R),\alpha \mid \alpha )=\mu $. This means that we can rewrite $%
V((p,R),d\mid \alpha )$ as follows:%
\begin{eqnarray}
V((p,R),d &\mid &\alpha )=\frac{d-\alpha }{1-\alpha }\cdot p_{R}(y=1\mid
a=1)+\frac{1-d}{1-\alpha }\cdot \mu  \label{V} \\
&=&\frac{\alpha -d}{\alpha }\cdot p_{R}(y=1\mid a=0)+\frac{d}{\alpha }\cdot
\mu
\end{eqnarray}%
It follows that the set of narratives that maximize $U$ for given $\left(
d,\alpha \right) $ only depends on the \textit{ordinal} ranking between $d$
and $\alpha $. Specifically, if $d>\alpha $, then $(p,R)$ should maximize $%
p_{R}(y=1\mid a=1)$; if $d<\alpha $, then $(p,R)$ should maximize $%
p_{R}(y=1\mid a=0)$; and if $d=\alpha $, then all feasible narratives induce 
$U=\mu -C(d-d^{\ast })$. Richness implies that there is $(p,R)$ such that
the slope of $V((p,R),d\mid \alpha )$ with respect to $d>\alpha $ is
strictly positive, and there is $(p,R)$ such that the slope of $%
V((p,R),d\mid \alpha )$ with respect to $d<\alpha $ is strictly negative.

It follows that the value function $\max_{(p,R)}V((p,R),d\mid \alpha )$ is
piecewise linear in $d$: It is linearly increasing (decreasing) in $d>\alpha 
$ ($d<\alpha $). Since $C$ is strictly convex, it follows that there is a
unique maximizer\ $d_{r}$ of $U((p,R),d\mid \alpha )$ in the range $d\geq
\alpha $, and a unique maximizer $d_{l}$ of $U((p,R),d\mid \alpha )$ in the
range of $d\leq \alpha $. In both cases, $\alpha $ cannot be the maximizer.
To see why, recall that $U((p,R),\alpha \mid \alpha )=\mu -C(\alpha -d^{\ast
})$ for any narrative $(p,R)$. We noted above that every $(s,d)\in
Supp(\sigma )$ must deliver $U(s,d)\geq \mu $. It follows that if $\alpha
\in \arg \max_{d}U((p,R),d\mid \alpha )$, then $\alpha =d^{\ast }$. But
since $C^{\prime }=0$ at $d=d^{\ast }$, it follows from (\ref{V}) that any
narrative $(p,R)$ with $p_{R}(y=1\mid a=0)>0$ satisfies $\max_{d>\alpha
}U((p,R),d\mid \alpha )>\mu $. Likewise, any narrative $(p,R)$ with $%
p_{R}(y=1\mid a=0)>0$ satisfies $\max_{d<\alpha }U((p,R),d\mid \alpha )>\mu $%
. We conclude that $d_{r}>\alpha $ and $d_{l}<\alpha $, and therefore the
support of the marginal of $\sigma $ over $d$ is weakly contained in $%
\{d_{l},d_{r}\}$. Because we have already established that this support
cannot be a singleton, the containment must be an identity.

It remains to establish that $d_{r}>d^{\ast }$ and $d_{l}<d^{\ast }$. Assume
the contrary such that without loss of generality, $d_{l}\geq d^{\ast }$.
Recall that $d_{l}$ is accompanied by a narrative $(p_{0},R_{0})$ for which $%
p_{R_{0}}(y=1\mid a=0)>0$. Therefore, the derivative of $U((p,R),d\mid
\alpha )$ with respect to $d$ is strictly negative at $d=d_{l}$, which means
that switching from $d_{l}$ to a slightly lower policy (without changing the
accompanying narrative) would generate a higher net anticipatory utility, a
contradiction.\medskip
\end{proof}

Thus, when the set of feasible narratives only involves perfect DAGs - yet
is sufficiently rich to enable belief distortion in either direction -
equilibrium must induce exactly two policies. Each of the two policies
deviates from the ideal point $d^{\ast }$ in a different direction. As the
proof of the result indicates, this polarization result does not directly
rely on the notion of narratives as causal models. Indeed, any model of
belief distortion that satisfies NSQD and richness would lead to the same
result. Causal models only play an indirect role in this sub-section:
Perfect DAGs imply NSQD and non-vacuousness of the richness property. They
will return to play a direct role in the next sub-section.

\subsection{Short Narratives}

In this sub-section we provide a complete equilibrium characterization for
the following specification. First, narratives must be short:\ They can
involve at most \textit{one} variable $x_{2}$ in addition to $a$ and $y$.
Second, $\mathcal{R}$ is the set of perfect DAGs with two or three nodes in
which $a$ is represented by an ancestral node. The only DAG in this class
that does $not$ induce $p_{R}(y\mid a)=\mu $ for all $a$ is the lever DAG $%
a\rightarrow x_{2}\rightarrow y$. Finally, $Q$ is large in the following
sense: There is an arbitrarily small constant $\delta >0$ such that for
every conditional distribution $(p(x_{2}\mid a,y))$, there is a conditional
distribution $q\in Q$ such that $\max_{a,y}\left\vert q(x_{2}=1\mid
a,y)-p(x_{2}=1\mid a,y)\right\vert <\delta $.

Our analysis in the previous sub-section implies that in any equilibrium $%
(\alpha ,\sigma )$, $Supp(\sigma )$ consists of two elements: a policy $%
d_{r}>d^{\ast }$ sustained by a lever narrative that employs some
distribution $q_{r}\in Q$, and a policy $d_{l}<d^{\ast }$ sustained by
another lever narrative that employs a different distribution $q_{l}\in Q$.
The following result refines this characterization.\medskip

\begin{proposition}
\label{prop_n=3}There is an essentially unique equilibrium $(\alpha ,\sigma
) $.\footnote{%
By essential uniqueness we mean that the definition of $q_{0}$ or $q_{1}$ is
unique up to relabeling of $x_{2}$.} In particular:\newline
(i) In the $\delta \rightarrow 0$ limit, $q_{r}$ is defined by $%
p(x_{2}=1\mid a,y)=y+a(1-y)$ and $q_{l}$ is defined by $p(x_{2}=1\mid
a,y)=y+(1-a)(1-y)$.\newline
(ii) $\alpha \in (\frac{1}{2},d^{\ast })$ when $d^{\ast }>\frac{1}{2}$, and $%
\alpha =\frac{1}{2}$ when $d^{\ast }=\frac{1}{2}$.
\end{proposition}

\begin{proof}
We established in the previous sub-section that $d_{r}$ is accompanied by a
narrative $(p,R)$ that maximizes $p_{R}(y=1\mid a=1)$; and likewise, $d_{l}$
is accompanied by a narrative $(p,R)$ that\ maximizes $p_{R}(y=1\mid a=0)$.
The only DAG that can induce non-constant $p_{R}(y\mid a)$ is $a\rightarrow
x_{2}\rightarrow y$. Therefore, the narratives that accompany both $d_{r}$
and $d_{l}$ involve this DAG, which we denote by $R$. To find the optimal
narrative that accompanies $d_{r}$, we need to find the quadruple $%
(p(x_{2}=1\mid a,y))_{a,y=0,1}$ that maximizes%
\begin{eqnarray*}
p_{R}(y &=&1\mid a=1)=\sum_{x_{2}}p(x_{2}\mid a=1)p(y=1\mid x_{2}) \\
&=&\sum_{x_{2}}\left( \sum_{y^{\prime }}p(y^{\prime })p(x_{2}\mid
a=1,y^{\prime })\right) \frac{\mu \sum_{a^{\prime }}p(a^{\prime
})p(x_{2}\mid a^{\prime },y=1)}{\sum_{y^{\prime \prime }}\sum_{a^{\prime
\prime }}p(a^{\prime \prime })p(y^{\prime \prime })p(x_{2}\mid a^{\prime
\prime },y^{\prime \prime })}
\end{eqnarray*}%
In the Appendix, we show that the solution in the $\delta \rightarrow 0$
limit is $p^{\ast }(x_{2}=1\mid a,y)=y+a(1-y)$, inducing%
\[
p_{R}^{\ast }(y=1\mid a=1)=\frac{\mu }{\mu +\alpha (1-\mu )} 
\]%
and, by NSQD,%
\[
p_{R}^{\ast }(y=1\mid a=0)=\frac{\mu ^{2}}{\mu +\alpha (1-\mu )} 
\]%
Therefore,%
\begin{eqnarray*}
V((p^{\ast },R),d &\mid &\alpha )=d\frac{\mu }{\mu +\alpha (1-\mu )}+(1-d)%
\frac{\mu ^{2}}{\mu +\alpha (1-\mu )} \\
&=&\mu +\frac{\mu (1-\mu )}{\mu +\alpha (1-\mu )}(d-\alpha )
\end{eqnarray*}

Likewise, the narrative that accompanies $d_{l}$ in the $\delta \rightarrow
0 $ limit involves the conditional distribution $p^{\ast \ast }(x_{2}=1\mid
a,y)=y+(1-a)(1-y)$, inducing%
\begin{eqnarray*}
p_{R}^{\ast \ast }(y &=&1\mid a=0)=\frac{\mu }{\mu +(1-\alpha )(1-\mu )} \\
p_{R}^{\ast \ast }(y &=&1\mid a=1)=\frac{\mu ^{2}}{\mu +(1-\mu )(1-\alpha )}
\end{eqnarray*}%
Therefore,%
\begin{eqnarray*}
V((p^{\ast \ast },R),d &\mid &\alpha )=d\frac{\mu ^{2}}{\mu +(1-\mu
)(1-\alpha )}+(1-d)\frac{\mu }{\mu +(1-\mu )(1-\alpha )} \\
&=&\mu -\frac{\mu (1-\mu )}{\mu +(1-\alpha )(1-\mu )}(d-\alpha )
\end{eqnarray*}

Denote%
\begin{eqnarray*}
U_{r}(\alpha ) &=&U((p^{\ast },R),d_{r}\mid \alpha )=V((p^{\ast
},R),d_{r}\mid \alpha )-C(d_{r}-d^{\ast }) \\
U_{l}(\alpha ) &=&U((p^{\ast \ast },R),d_{l}\mid \alpha )=V((p^{\ast \ast
},R),d_{l}\mid \alpha )-C(d_{l}-d^{\ast })
\end{eqnarray*}%
Denote $\Delta =\left\vert d-\alpha \right\vert $, $e=\alpha -d^{\ast }$.
Then, we can write%
\begin{eqnarray}
U_{r}(\alpha ) &=&\max_{\Delta \leq 1-\varepsilon -\alpha }\left[ \mu +\frac{%
\mu (1-\mu )}{\mu +\alpha (1-\mu )}\Delta -C(\Delta +e)\right]  \label{UrUl}
\\
U_{l}(\alpha ) &=&\max_{\Delta \leq \alpha -\varepsilon }\left[ \mu +\frac{%
\mu (1-\mu )}{\mu +(1-\alpha )(1-\mu )}\Delta -C(\Delta -e)\right]  \nonumber
\end{eqnarray}%
Recall that by assumption, $d^{\ast }\geq \frac{1}{2}$. Suppose $\alpha
>d^{\ast }$. Then, $\alpha >\frac{1}{2}$ and $e>0$. It is then clear from (%
\ref{UrUl}) that $U_{r}(\alpha )<U_{l}(\alpha )$, contradicting equilibrium.
Now suppose $\alpha <\frac{1}{2}$. Then, $e<0$, and it is clear from (\ref%
{UrUl}) that $U_{r}(\alpha )>U_{l}(\alpha )$, again contradicting
equilibrium. It follows that $\alpha \in \lbrack \frac{1}{2},d^{\ast }]$.
Furthermore, since $U_{r}(\alpha )$ is strictly decreasing in $\alpha $
while $U_{l}(\alpha )$ is strictly increasing in $\alpha $, there is at most
one value of $\alpha $ for which $U_{r}(\alpha )=U_{l}(\alpha )$, hence
equilibrium must be unique.\medskip
\end{proof}

The characterization has a number of noteworthy properties. First, the lever
narrative that sustains either of the two equilibrium policies selects the
intermediate variable $x_{2}$ such that it is highly correlated with both
the desired outcome $y=1$ and the advocated policy. Specifically, the
selected variable is such that one particular value is attained whenever $%
y=1 $ \textit{or} the favored action is taken.

For illustration, recall the US trade policy debate described in the
Introduction. In this context, our characterization approximates the
following prevailing narratives. The lever narrative that sustains a policy
with a protectionist bias (relative to the agent's ideal point) will involve
a variable like \textquotedblleft imports from China\textquotedblright ,
because low imports are associated with trade restrictions as well as with
high employment in the local manufacturing sector, even if the latter
correlation is not causal but due to a confounding factor (such as exogenous
technology changes that affect outsourcing of production). Likewise, the
lever narrative that sustains a trade policy with a liberalized bias will
select a variable like "industrial exports".

Second, the anticipatory utility induced by the equilibrium narratives
exhibits a diminishing-returns property. That is, when $\alpha $ increases
(decreases), the narrative that advocates right-leaning (left-leaning)
policies has lower anticipatory value. This property is intuitive:
narratives generate false hopes about counterfactual policies; as the
historical action frequency leans in the same direction as the narrative,
the ability to sell this illusion diminishes. In turn, the
diminishing-returns property implies two features of equilibrium: essential
uniqueness (specifically, the marginal equilibrium distribution over
policies is unique) and a \textquotedblleft centrist bias\textquotedblright\
(i.e., the historical action frequency lies between $\frac{1}{2}$ and $%
d^{\ast }$.

\section{Opportunity Narratives}

Our analysis in the previous section ruled out imperfect DAGs, which include
the opportunity narrative we encountered in Section 3. In this section we
explore the implication of allowing for imperfect DAGs. We focus our
analysis on the case in which only a single auxiliary variable can be used
(i.e., $n=3$). Thus, the set of feasible DAGs is the set of all DAGs with up
to three nodes, in which $a$ is represented by an ancestral node. The only
imperfect DAG in this class is $a\rightarrow x_{2}\leftarrow y$. We assume
throughout that $d^{\ast }=\frac{1}{2}$.

The following result establishes a polarization result akin to that of
Section 4.2.\medskip

\begin{proposition}
If $(Q,\mathcal{R})$ is rich in the sense of Section 4.2, then any
equilibrium assigns positive probability to at least one policy $d>d^{\ast }$
and one policy $d<d^{\ast }$.
\end{proposition}

\begin{proof}
Assume the contrary - without loss of generality, there is an equilibrium $%
(\alpha ,\sigma )$ that assigns probability one to policies $d\geq d^{\ast }=%
\frac{1}{2}$. Therefore, $\alpha \geq \frac{1}{2}$. If the DAG $a\rightarrow
y\leftarrow x_{2}$ is never played in this equilibrium, we are back with the
model of Section 4.2, where this possibility was ruled out.

Now suppose that $Supp(\sigma )$ includes a narrative-policy pair $((p,R),d)$
in which $R:a\rightarrow y\leftarrow x_{2}$. Let us first establish that,
for one such pair, $p_{R}(y=1\mid a=1)\neq p_{R}(y=1\mid a=0)$ . Assume the
contrary for every such $(p,R)$. This means that if we switched to the DAG $%
R^{\prime }:y\leftarrow x_{2}$, we would have $p_{R^{\prime }}(y)=p_{R}(y)$.
However, since $p_{R^{\prime }}(y)\equiv p(y)$, we have $p_{R}(y=1\mid
a)=\mu $ for all $a$. This means that the narrative-policy pair $((p,R),d)$
induces the same net anticipatory utility as if the narrative involved the
DAG $a\rightarrow y$. Since we can perform this substitution for every
narrative-policy pair in $Supp(\sigma )$ that involves the DAG $a\rightarrow
y\leftarrow x_{2}$, we are back in the case of Section 4.2, which again
leads to a contradiction.

From now on, assume without loss of generality that for every
narrative-policy pair $((p,R),d)$ in which $R:a\rightarrow y\leftarrow x_{2}$%
, $p_{R}(y=1\mid a=1)\neq p_{R}(y=1\mid a=0)$. Suppose $d=d^{\ast }$. Since $%
C$ is flat at this point, a deviation to the narrative policy pair $%
((p,R),d^{\prime })$, where $d^{\prime }$ is slightly different from $%
d^{\ast }$ in the direction of the action $a$ that has the higher $%
p_{R}(y=1\mid a)$ would generate higher net anticipatory utility,
contradicting the definition of equilibrium. Therefore, $d>d^{\ast }$. In
particular, this means that $\alpha >\frac{1}{2}$. If $p_{R}(y=1\mid
a=1)<p_{R}(y=1\mid a=0)$, a switch to the narrative-policy pair $((p,R),1-d)$
would increase gross anticipatory utility without changing $C$, a
contradiction.

Thus, $\alpha >\frac{1}{2}$ and $Supp(\sigma )$ includes a narrative-policy
pair $((p,R),d)$ in which $R:a\rightarrow y\leftarrow x_{2}$, $d>\frac{1}{2}$
and $p_{R}(y=1\mid a=1)>p_{R}(y=1\mid a=0)$. Write down the explicit formula
for $p_{R}(y\mid a)$:%
\begin{eqnarray}
p_{R}(y &=&1\mid a)=\sum_{x_{2}}p(x_{2})p(y=1\mid a,x_{2})  \label{p_R(y|a)}
\\
&=&\sum_{x_{2}}\left( \sum_{a^{\prime \prime }}p(a^{\prime \prime
})\dsum\limits_{y^{\prime \prime }}p(y^{\prime \prime })p(x_{2}\mid
a^{\prime \prime },y^{\prime \prime })\right) \frac{p(a)(p(y=1))p(x_{2}\mid
a,y=1)}{\dsum\limits_{y^{\prime }}p(y^{\prime })p(a)p(x_{2}\mid a,y^{\prime
})}  \nonumber \\
&=&\mu \sum_{x_{2}}\frac{p(x_{2}\mid a,y=1)}{\dsum\limits_{y^{\prime
}}p(y^{\prime })p(x_{2}\mid a,y^{\prime })}\sum_{a^{\prime \prime
}}p(a^{\prime \prime })\dsum\limits_{y^{\prime \prime }}p(y^{\prime \prime
})p(x_{2}\mid a^{\prime \prime },y^{\prime \prime })  \nonumber
\end{eqnarray}%
For $a=1$, this expression becomes%
\begin{eqnarray*}
\mu \sum_{x_{2}}p(x_{2} &\mid &a=1,y=1)\frac{\alpha
\dsum\nolimits_{y}p(y)p(x_{2}\mid a=1,y)+(1-\alpha
)\dsum\nolimits_{y}p(y)p(x_{2}\mid a=0,y)}{\dsum\nolimits_{y}p(y)p(x_{2}\mid
a=1,y)} \\
&=&\mu \sum_{x_{2}}p(x_{2}\mid a=1,y=1)\left[ \alpha +(1-\alpha )\frac{%
\dsum\nolimits_{y}p(y)p(x_{2}\mid a=0,y)}{\dsum\nolimits_{y}p(y)p(x_{2}\mid
a=1,y)}\right] \\
&=&\mu \left[ \alpha +(1-\alpha )\sum_{x_{2}}p(x_{2}\mid a=1,y=1)\frac{%
\dsum\nolimits_{y}p(y)p(x_{2}\mid a=0,y)}{\dsum\nolimits_{y}p(y)p(x_{2}\mid
a=1,y)}\right]
\end{eqnarray*}

Likewise, for $a=0$, (\ref{p_R(y|a)}) becomes%
\[
\mu \left[ (1-\alpha )+\alpha \sum_{x_{2}}p(x_{2}\mid a=0,y=1)\frac{%
\dsum\nolimits_{y}p(y)p(x_{2}\mid a=1,y)}{\dsum\nolimits_{y}p(y)p(x_{2}\mid
a=0,y)}\right] 
\]%
Denote%
\begin{eqnarray*}
A &=&\sum_{x_{2}}p(x_{2}\mid a=1,y=1)\frac{\dsum\nolimits_{y}p(y)p(x_{2}\mid
a=0,y)}{\dsum\nolimits_{y}p(y)p(x_{2}\mid a=1,y)} \\
B &=&\sum_{x_{2}}p(x_{2}\mid a=0,y=1)\frac{\dsum\nolimits_{y}p(y)p(x_{2}\mid
a=1,y)}{\dsum\nolimits_{y}p(y)p(x_{2}\mid a=0,y)}
\end{eqnarray*}%
Since $p_{R}(y=1\mid a=1)>p_{R}(y=1\mid a=0)$, $A>B$. And since $%
p_{R}(y=1\mid a=1)>\mu $, $A>1$. The net anticipatory utility generated by $%
((p,R),\alpha )$ can thus be written as%
\begin{eqnarray}
d\cdot p_{R}(y &=&1\mid a=1)+(1-d)\cdot p_{R}(y=1\mid a=0)-C(d-\frac{1}{2})
\label{V_beforedeviation} \\
&=&\mu \left[ d(\alpha +(1-\alpha )A)+(1-d)((1-\alpha )+\alpha B)\right]
-C(d-\frac{1}{2})  \nonumber
\end{eqnarray}%
Now consider a deviation to the narrative-policy pair $((\tilde{p},R),1-d)$,
where $\tilde{p}$ is defined by%
\[
\tilde{p}(x_{2}\mid a,y)\equiv p(x_{2}\mid 1-a,y) 
\]%
That is, $\tilde{p}$ is a mirror image of $p$. By assumption, $\tilde{p}$ is
feasible. Define $\tilde{A}$ and $\tilde{B}$ accordingly. By construction, $%
\tilde{A}=B$ and $\tilde{B}=A$. Therefore, the net anticipatory utility
generated by $((\tilde{p},R),1-d)$ is%
\begin{eqnarray*}
(1-d)\cdot \tilde{p}_{R}(y &=&1\mid a=1)+d\cdot \tilde{p}_{R}(y\mid
a=0)-C((1-d)-\frac{1}{2}) \\
&=&\mu \left[ (1-d)(\alpha +(1-\alpha )B)+d((1-\alpha )+\alpha A)\right] -C(%
\frac{1}{2}-d)
\end{eqnarray*}%
Since $d,\alpha >\frac{1}{2}$ and $A>1$, this expression exceeds (\ref%
{V_beforedeviation}), a contradiction.\medskip
\end{proof}

Unlike the case of perfect DAGs, the DAG $a\rightarrow x_{2}\leftarrow y$
does not satisfy the NSQD property, and therefore the proof resorts to other
arguments. The key question is whether, assuming all equilibrium policies
lie on one side of $d^{\ast }=\frac{1}{2}$, a narrative-policy pair $%
((p,a\rightarrow x_{2}\leftarrow y),d)\in Supp(\sigma )$ can be destabilized
by a deviation to a \textquotedblleft mirror\textquotedblright\ pair. The
answer is not obvious, and our proof relies on the particular structure of
the imperfect three-node DAG $a\rightarrow x_{2}\leftarrow y$.

The result is weaker than its analogue in Section 4.2. In particular, we are
unable to determine whether equilibrium will sustain $exactly$ one policy on
each side of $d^{\ast }$ for general cost functions. However, when costs are
sufficiently small, we obtain a stronger characterization.\medskip

\begin{proposition}
\label{prop_opportunity}Suppose (as in Section 4.3) that there is an
arbitrarily small constant $\delta >0$ such that for every conditional
distribution $(p(x_{2}\mid a,y))$ there is $q\in Q$ such that $%
\max_{a,y}\left\vert q(x_{2}=1\mid a,y)-p(x_{2}=1\mid a,y)\right\vert
<\delta $. Then, if $C^{\prime }(\cdot )$ and $\varepsilon $ are
sufficiently small, there is a unique equilibrium, in which $\alpha =\frac{1%
}{2}\ $and $Supp(\sigma )$ consists of:\newline
(i) An opportunity narrative that consists of the DAG $a\rightarrow
y\leftarrow x_{2}$ and the conditional distribution $p(x_{2}=1\mid
a,y)\approx y+(1-a)(1-y)$, coupled with a policy $d_{r}\approx 1$.\newline
(ii) An opportunity narrative that consists of the DAG $a\rightarrow
y\leftarrow x_{2}$ and the conditional distribution $p(x_{2}=1\mid
a,y)\approx y+a(1-y)$, coupled with a policy $d_{l}\approx 0$.\footnote{%
If $d^{\ast }>\frac{1}{2}$, a similar result holds, where the only
difference is that $\alpha \in (\frac{1}{2},d^{\ast }).$}\medskip
\end{proposition}

\begin{proof}
In Section 4.3, we derived, for each $a=0,1$, a lever narrative that
sustains $p_{R}(y=1\mid a)-p_{R}(y=1\mid 1-a)>0$ for any given $\alpha \in
(0,1)$. Since this difference is the derivative of $V$ with respect to $d$,
it follows that if $C^{\prime }$ is sufficiently small, the only policies
that survive in equilibrium are the extreme points $d=1-\varepsilon $ and $%
d=\varepsilon $. It follows that in order to characterize equilibrium in the
low $\varepsilon $ limit, we only need to look for the narratives $(p,R)$
that maximize $p_{R}(y=1\mid a)$ for each $a=0,1$.

In Section 4.3, we saw that the largest $p_{R}(y=1\mid a=1)$ and $%
p_{R}(y=1\mid a=0)$ that lever narratives can attain are $\mu /[\mu +(1-\mu
)\alpha ]$\ and $\mu /[\mu +(1-\mu )(1-\alpha )]$, respectively. In the
Appendix, we show that the largest $p_{R}(y=1\mid a=1)$ and $p_{R}(y=1\mid
a=0)$ that opportunity narratives can attain are $1-\alpha (1-\mu )$ and $%
1-(1-\alpha )(1-\mu )$, respectively. A simple calculation establishes that 
\[
1-\alpha (1-\mu )>\frac{\mu }{\mu +(1-\mu )\alpha } 
\]%
for any $\alpha \in (0,1)$. It follows that the prevailing narrative-policy
pairs in any equilibrium in the $\varepsilon ,\delta \rightarrow 0$ limit
are as described in the statement of the proposition. In equilibrium, these
pairs must deliver the same net anticipatory utility:%
\[
1-\alpha (1-\mu )-C(1-\frac{1}{2})=1-(1-\alpha )(1-\mu )-C(-\frac{1}{2}) 
\]%
which holds if and only if $\alpha =\frac{1}{2}$.\medskip
\end{proof}

Thus, when the set of feasible three-node DAGs is unrestricted, the set $Q$
is rich and the cost $C$ is low, the narratives that prevail in equilibrium
are opportunity narratives and they sustain extreme policies. Surprisingly,
the opportunity narrative that sustains an extreme right (left) policy
employs the $same$ third variable that was employed by the equilibrium lever
narrative that sustained the extreme left (right) in Section 4.3. We saw an
inkling of this effect in the illustrative example of Section 3: The same
variable can feature in narratives that support radically different
policies; what changes is the role that this variable plays in the
narrative's causal structure.

\section{Conclusion}

The model presented in this paper formalized a number of intuitions
regarding the role of narratives in the formation of popular political
opinions. Our model was based on two main ideas.\medskip

\noindent \textit{What are narratives and how do they shape beliefs?} In our
model, narratives are formalized as causal models (represented by DAGs) that
describe how actions map into consequences. Different narratives employ
different intermediate variables and arrange them differently in the causal
scheme. Narratives shape beliefs in the sense that beliefs emerge from
fitting causal models to long-run correlations between the variables that
appear in the narrative. These beliefs are used to evaluate policies.\medskip

\noindent \textit{How does the public select between competing narratives?}
Our behavioral assumption was that in the presence of conflicting
narrative-policy pairs, the public (a representative agent in this paper)
selects between them \textquotedblleft hedonically\textquotedblright\ -
i.e., according to the anticipatory utility induced by each of these pairs.
This is consistent with the basic intuition that people are drawn to
\textquotedblleft hopeful\textquotedblright\ stories.\medskip

The main insights that emerged as results of our formalism can be summarized
as follows. First, narratives are employed to \textquotedblleft sell false
hopes\textquotedblright : They involve misspecified causal models that
generate biased beliefs regarding the consequences of counterfactual
policies. Second, the same variable can serve two conflicting narratives
with a different causal structure (e.g., \textquotedblleft lever
narrative\textquotedblright\ vs. \textquotedblleft opportunity
narrative\textquotedblright ) in the service of conflicting policies. Third,
multiplicity of dominant narrative-policy pairs can be a fundamental
property of long-run equilibrium in the \textquotedblleft battle over public
opinion\textquotedblright . Indeed, growing popularity of one policy can
strengthen the appeal of a narrative that supports an opposing policy. This
\textquotedblleft diminishing returns\textquotedblright\ property leads to
additional properties of equilibrium (uniqueness, centrist bias) in specific
settings. Finally, when we rule out narratives that convey false beliefs
regarding the status quo, linear narratives are without loss of generality.

Our analysis leaves a number of open technical problems. First, Section 4.3
provided a complete equilibrium characterization for perfect DAGs and rich $%
Q $ in the case of $n=3$. We also know that for $n=4$, equilibrium
narratives have the longer linear form $a\rightarrow x_{2}\rightarrow
x_{3}\rightarrow y $. Naturally, we conjecture that for general $n$,
prevailing narratives are linear chains of length $n$. But what are the
conditional beliefs over consequences that these prevailing narratives
induce? Finally, the case of general $n$ and an unrestricted set of feasible
DAGs (including imperfect ones) is almost entirely open; the only analysis
we have been able to carry out for this domain is the $n=3$ example of
Section 4. A broad question that is common to these two cases is whether our
definition of equilibrium generates a force that favors narratives that
involve many variables.

\noindent {\LARGE Appendix: Proofs}\bigskip

\noindent \textbf{Proof of Proposition \ref{existence}}

\noindent Consider an auxiliary two-player game. Player 1's strategy space
is $D$, and $\alpha $ denotes an element in this space. Player 2's strategy
space is $\Delta (Q\times \mathcal{R}\times D)$, and $\beta $ denotes an
element in this space. Observe that when we fix $\alpha $ and $\mu $, an
element $q\in Q$ induces unambiguously an element $p_{q}\in P_{\alpha ,\mu }$%
.

The payoff of player $1$ from the strategy profile $(\alpha ,\beta )$ is%
\[
\sum_{(q,R,d)}\beta (q,R,d)U((p_{q},R),d\mid \alpha ) 
\]%
Note that since $p_{R}$ is a continuous function of $\alpha $, so is $U$.
The payoff of player $2$ from $(\sigma ,\alpha )$ is%
\[
-\left( \alpha -\dsum\nolimits_{(q,R,d)}\beta ((p_{q},R),d)d\right) ^{2} 
\]%
A Nash equilibrium in this auxiliary game is equivalent to our notion of
equilibrium. The strategy spaces and payoff functions of the two players in
the auxiliary game satisfy standard conditions for the existence of Nash
equilibrium.\bigskip

\noindent \textbf{Proof of Proposition \ref{prop_linearDAG}}

\noindent The proof proceeds in the three main steps.\medskip

\noindent \textit{Step 1: Deriving an auxiliary \textquotedblleft clique
factorization\textquotedblright\ formula}

\noindent Consider a non-linear perfect DAG $(N,R)$, where $N=\{1,...,n\}$, $%
n>2$. We say that a subset of nodes $C\subseteq N$ is a $clique$ if for
every $i,j\in C$, $iRj$ or $jRi$. We say that a clique is $maximal$ if it is
not contained in another clique. Let $\mathcal{C}$ be the collection of
maximal cliques in the DAG.

The following is standard material in the Bayesian-Networks literature.
Because $(N,R)$ is perfect, we can construct an auxiliary (non-directed) $%
tree$ whose set of nodes is $\mathcal{C}$, such that for every pair of nodes 
$C$ and $C^{\prime }$ in this tree, $C\cap C^{\prime }$ is contained in any $%
C^{\prime \prime }$ that lies along the path that connects $C$ and $%
C^{\prime }$ (the path is unique, by the definition of a tree). Such a tree
is referred to in the literature as a \textit{junction tree}. Given a
junction tree, we say that $S\subseteq N$ is a \textit{separator} if there
are two adjacent tree nodes $C$ and $C^{\prime }$ such that $S=C\cap
C^{\prime }$. Let $\mathcal{S}$ be the set of separators for a given
junction tree constructed from $\mathcal{C}$. Then, for any distribution $%
p^{\prime }\in \Delta (X)$ with full support that is consistent with $(N,R$)
(i.e., in the sense that $p_{R}=p$),%
\[
p^{\prime }(x)=\frac{\dprod\nolimits_{C\in \mathcal{C}}p^{\prime }(x_{C})}{%
\dprod\nolimits_{S\in \mathcal{S}}p^{\prime }(x_{S})} 
\]%
For an exposition of these results, see Cowell et al. (1999), pp. 52-69.

Now, our objective distribution $p$ is $not$ necessarily consistent with $R$%
. However, $p_{R}$ is consistent with $R$ by definition. Furthermore, a key
feature of perfect DAGs is that they do not distort the marginal
distributions over cliques - i.e., $p_{R}(x_{C})\equiv p(x_{C})$ for every $%
C\in \mathcal{C}$ (see Spiegler (2017) for further details). It follows that
for every objective distribution $p$ and a perfect DAG $(N,R)$, we can write%
\begin{equation}
p_{R}(x)\equiv \frac{\dprod\nolimits_{C\in \mathcal{C}}p(x_{C})}{%
\dprod\nolimits_{S\in \mathcal{S}}p(x_{S})}  \label{junctiontreeformula}
\end{equation}%
where $\mathcal{C}$ is the set of maximal cliques in $(N,R)$ and $\mathcal{S}
$ is the set of separators in some junction tree constructed out of $%
\mathcal{C}$.

Let $C_{1},C_{m}\in \mathcal{C}$ be two cliques in $(N,R)$ that include the
nodes $1$ and $n$, respectively. Furthermore, for a given junction tree
representation of the DAG, select these cliques to be minimally distant from
each other - i.e., $1,n\notin C$ for every $C$ along the junction-tree path
between $C_{1}$ and $C_{m}$.

If $C_{1}=C_{m}$, then by our earlier observation that perfect DAGs do not
distort the marginals of collections of variables that form a clique, it
follows that $p_{R}(x_{1},x_{n})\equiv p(x_{1},x_{n})$ and therefore $%
p_{R}(x_{n}\mid x_{1})\equiv p(x_{n}\mid x_{1})$ - i.e. we can replace the
original DAG with the degenerate linear DAG $1\rightarrow n\ $and obtain the
same subjective conditional distribution over $x_{n}$. The same deviation
holds if there is $no$ junction-tree path between $C_{1}$ and $C_{m}$,
because this means that $x_{1}\perp x_{n}$ according to $p_{R}$, and
therefore $p_{R}(x_{n}\mid x_{1})\equiv p(x_{n}\mid x_{1})$.

Thus, from now on, assume that $C_{1}\neq C_{m}$ and there is a
junction-tree path between $C_{1}$ and $C_{m}$. Enumerate all the nodes in
the junction tree and turn it into a directed tree, such that $C_{1}$ is its
root node. For every $k=2,...,\left\vert \mathcal{C}\right\vert $, let $pa(k)
$ denote the index of the direct parent of $C_{k}$ - i.e. the junction tree
has a direct link $C_{pa(k)}\rightarrow C_{k}$. In particular, let $%
C_{1},C_{2},...,C_{m}$ be the tree nodes along the path between $C_{1}$ and $%
C_{m}$, such that this path is $C_{1}\rightarrow C_{2}\rightarrow \cdots
\rightarrow C_{m}$. By the definition of a junction tree, if $i\in
C_{k},C_{j}$ for some $1\leq k<j\leq m$, then $i\in C_{h}$ for every $%
h=k+1,...,j-1$. And since the cliques $C_{1},...,C_{m}$ are maximal, it
follows that every $C_{k}$ along the sequence $C_{0},...,C_{m+1}$ must
introduce at least one element $i\notin \cup _{j<k}C_{j}$. As a result, it
must be the case that $m\leq n-1$.

Now, repeatedly apply the identity%
\[
p(x_{C_{k}})=p(x_{C_{k}\cap C_{pa(k)})})p(x_{C_{k}-C_{pa(k)}}\mid
x_{C_{k}\cap C_{pa(k)}}) 
\]%
to (\ref{junctiontreeformula}) for every $k\geq 2$, and obtain the following
equivalent formula:%
\[
p_{R}(x)\equiv p(x_{C_{1}})\cdot \dprod\nolimits_{k=2}^{\left\vert \mathcal{C%
}\right\vert }p(x_{C_{k}-C_{pa(k)}}\mid x_{C_{k}\cap C_{pa(k)}}) 
\]%
Furthermore, by the definition of the junction tree, for every $k>m$, $%
C_{k}-C_{pa(k)}$ and $C^{\ast }=C_{1}\cup \cdots \cup C_{m}$ are mutually
disjoint. Therefore,%
\begin{equation}
p_{R}(x_{C^{\ast }})\equiv
p(x_{C_{1}})\dprod\nolimits_{k=2}^{m}p(x_{C_{k}-C_{k-1}}\mid x_{C_{k}\cap
C_{k-1}})  \label{clique_factorization}
\end{equation}%
\medskip

\noindent \textit{Step 2: Obtaining a linear-DAG factorization}

\noindent We begin this step by deriving the subjective conditional
probability $p_{R}(x_{n}\mid x_{1})$ from (\ref{clique_factorization}).
Recall that from the definition of $C_{1}$ and $C_{m}$ it follows that $1\in
C_{1}$, $n\in C_{m}$, and $1,n\notin C_{k}$ for every $k=2,...,m-1$. Denote $%
C_{0}=\{1\}$ and observe that $p(x_{C_{1}})=p(x_{1})p(x_{C_{1}-\{1\}}\mid
x_{1})$. Then,%
\begin{equation}
p_{R}(x_{n}\mid x_{1})=\sum_{x_{C^{\ast
}-\{1,n\}}}\dprod\nolimits_{k=1}^{m}p(x_{C_{k}-C_{k-1}}\mid x_{C_{k}\cap
C_{k-1}})  \label{conditional_factorization}
\end{equation}%
We can draw an immediate conclusion from this formula. Suppose that there is
some $i\in C^{\ast }-\{1,n\}$ such that $i\in C_{k}$ for a \textit{unique} $%
k=1,...,m$. Then, the variable $x_{i}$ appears in only one term in (\ref%
{conditional_factorization}), namely $p(x_{C_{k}-C_{k-1}}\mid x_{C_{k}\cap
C_{k-1}})$. Moreover, by assumption, $i\in C_{k}-C_{k-1}$. Therefore, we can
rewrite this term as follows:%
\[
p(x_{C_{k}-C_{k-1}}\mid x_{C_{k}\cap C_{k-1}})=p(x_{C_{k}-(C_{k-1}\cup
\{i\})}\mid x_{C_{k}\cap C_{k-1}})p(x_{i}\mid x_{(C_{k}\cup C_{k-1})-\{i\}}) 
\]%
This means we can rewrite $p_{R}(x_{n}\mid x_{1})$ as follows:%
\[
\sum_{x_{C^{\ast }-\{1,n\}}}\dprod\nolimits_{h\neq k}p(x_{C_{h}-C_{h-1}}\mid
x_{C_{h}\cap C_{h-1}})p(x_{C_{k}-(C_{k-1}\cup \{i\})}\mid x_{C_{k}\cap
C_{k-1}})p(x_{i}\mid x_{(C_{k}\cup C_{k-1})-\{i\}})= 
\]%
\[
\sum_{x_{C^{\ast }-\{1,n,i\}}}\dprod\nolimits_{h\neq
k}p(x_{C_{h}-C_{h-1}}\mid x_{C_{h}\cap C_{h-1}})p(x_{C_{k}-(C_{k-1}\cup
\{i\})}\mid x_{C_{k}\cap C_{k-1}})\sum_{x_{i}}p(x_{i}\mid x_{(C_{k}\cup
C_{k-1})-\{i\}})= 
\]%
\[
\sum_{x_{C^{\ast }-\{1,n,i\}}}\dprod\nolimits_{h\neq
k}p(x_{C_{h}-C_{h-1}}\mid x_{C_{h}\cap C_{h-1}})p(x_{C_{k}-(C_{k-1}\cup
\{i\})}\mid x_{C_{k}\cap C_{k-1}}) 
\]

This is the same formula we would have if we removed $i$ (and the links
associated with this node) from the original DAG in the first place.
Therefore, without loss of generality, we can assume that every $i\in
C^{\ast }-\{1,n\}$ belongs to at least two cliques $C_{k}$, $k=1,...,m$.
Furthermore, by the definition of a junction tree, these two cliques are
consecutive, $C_{k}$ and $C_{k+1}$. In particular, this means that $%
C_{1}-C_{2}=\{1\}$, $C_{m}-C_{m-1}=\{n\}$, and $C_{k}-C_{k-1}\subseteq
C_{k+1}\cap C_{k}$ for every $k=1,...,m-1$. The latter observation implies
that for every $k=1,...,m-1$, $(C_{k+1}\cap C_{k})-(C_{k}-C_{k-1})$ is
weakly contained in $C_{k}\cap C_{k-1}$. Therefore, $p(x_{C_{k}-C_{k-1}}\mid
x_{C_{k}\cap C_{k-1}})=p(x_{C_{k+1}\cap C_{k}}\mid x_{C_{k}\cap C_{k-1}})$,
such that we can replace the term $p(x_{C_{k}-C_{k-1}}\mid x_{C_{k}\cap
C_{k-1}})$ in (\ref{clique_factorization}) with the equivalent term $%
p(x_{C_{k+1}\cap C_{k}}\mid x_{C_{k}\cap C_{k-1}})$. Finally, perform
another change in (\ref{clique_factorization}), by replacing $p(x_{C_{1}})$
with the equivalent term $p(x_{1})p(x_{C_{2}\cap C_{1}}\mid x_{1})$. After
these changes are performed, (\ref{clique_factorization}) is transformed
into a Bayesian-network factorization formula with respect to a linear DAG%
\[
1\rightarrow (C_{2}\cap C_{1})\rightarrow (C_{3}\cap C_{2})\cdots
\rightarrow (C_{m}\cap C_{m-1})\rightarrow m 
\]%
This DAG has at most $m+1\leq n$ nodes.\medskip

\noindent \textit{Step 3: Transforming the intermediate linear-DAG nodes\
into binary variables}

\noindent\ For every $k=2,...,m-1$, define $z_{k}=x_{C_{k}\cap C_{k-1}}$,
and let $z_{k}^{\ast }$ be one arbitrary value that the variable $z_{k}$ can
get. (Because $p$ has full support, at least two values of each $z_{k}$ have
positive probability.) Observe that%
\[
p_{R}(y|a)=\sum_{z_{2},...,z_{m-1}}p(z_{2}|a)p(z_{3}|z_{2})\cdots
p(z_{m-1}|z_{m-2})p(y|z_{m-1}) 
\]%
is equal to%
\[
\dsum\limits_{z_{2},...,z_{k-1}}p(z_{2}|a)\cdots
p(z_{k-1}|z_{k-2})\dsum\limits_{z_{k+1}}\left(
\dsum\limits_{z_{k}}p(z_{k}|z_{k-1})p(z_{k+1}|z_{k})\right) \cdots
\dsum\limits_{z_{m-1}}p(z_{m-1}|z_{m-2})p(y|z_{m-1}) 
\]%
The expression in the large parenthesis can be written as%
\[
p(z_{k}=z_{k}^{\ast }|z_{k-1})p(z_{k+1}|z_{k}=z_{k}^{\ast })+p(z_{k}\neq
z_{k}^{\ast }|z_{k-1})p(z_{k+1}|z_{k}\neq z_{k}^{\ast }) 
\]%
This is the only place in the\ formula for $p_{R}(y|a)$ where $z_{k}$ makes
an appearance. Therefore, without loss of generality, we can transform $%
z_{k} $ into a binary variable that gets the value $1$ when $%
z_{k}=z_{k}^{\ast }$ and the value $0$ when $z_{k}\neq z_{k}^{\ast }$. The
distribution $p^{\prime }$ over $a$, $y$ and the other $m-2$ binary
variables is thus derived from $p$ via the above series of steps. The
requirement that $p^{\prime }$ has full support is therefore satisfied
because $z_{k}$ gets at least two values.\bigskip

\noindent \textbf{Missing step in the proof of Proposition \ref{prop_n=3}}

\noindent \textit{Let }$R^{L}:a\rightarrow x_{2}\rightarrow y.$\textit{\ }%
Our objective is to show that%
\begin{eqnarray*}
p_{R^{L}}(y &=&1|a=1)\leq \frac{\mu }{\mu +\alpha (1-\mu )} \\
p_{R^{L}}(y &=&1|a=0)\leq \frac{\mu }{\mu +(1-\alpha )(1-\mu )}
\end{eqnarray*}%
in the $\delta \rightarrow 0$ limit. To derive these upper bounds, note
first that%
\[
p_{R^{L}}(y=1|a=1)=\sum_{x_{2}=0,1}p(x_{2}|a=1)p(y=1|x_{2}) 
\]%
Using the notation $p_{ay}\equiv p(x_{2}=1|a,y)$, $p_{R^{L}}(y=1|a=1)$ can
be rewritten as%
\begin{eqnarray*}
&&[\mu p_{11}+(1-\mu )p_{10}]\frac{\mu \lbrack \alpha p_{11}+(1-\alpha
)p_{01}]}{(1-\mu )[\alpha p_{10}+(1-\alpha )p_{00}]+\mu \lbrack \alpha
p_{11}+(1-\alpha )p_{01}]} \\
&&+[1-\mu p_{11}-(1-\mu )p_{10}]\frac{\mu \lbrack 1-\alpha p_{11}-(1-\alpha
)p_{10}]}{(1-\mu )[1-\alpha p_{10}-(1-\alpha )p_{00}]+\mu \lbrack 1-\alpha
p_{11}-(1-\alpha )p_{01}]}
\end{eqnarray*}%
This expression is a convex combination of two expressions,%
\begin{equation}
\frac{\mu \lbrack \alpha p_{11}+(1-\alpha )p_{01}]}{(1-\mu )[\alpha
p_{10}+(1-\alpha )p_{00}]+\mu \lbrack \alpha p_{11}+(1-\alpha )p_{01}]}
\label{C}
\end{equation}%
and%
\begin{equation}
\frac{\mu \lbrack 1-\alpha p_{11}-(1-\alpha )p_{10}]}{(1-\mu )[1-\alpha
p_{10}-(1-\alpha )p_{00}]+\mu \lbrack 1-\alpha p_{11}-(1-\alpha )p_{01}]}
\label{D}
\end{equation}%
Suppose (\ref{C}) is greater or equal to (\ref{D}). Then $p_{R^{L}}(y=1|a=1)$
attains a maximum only if $p_{10}=p_{11}=1.$ Given this, (\ref{C}) attains a
maximum at $p_{01}=1$ and $p_{00}=0$. At these values,%
\[
p_{R^{L}}(y=1|a=1)=\frac{\mu }{\mu +\alpha (1-\mu )} 
\]%
and indeed, (\ref{C}) is greater than (\ref{D}).

Using analogous arguments,%
\[
p_{R^{L}}(y=1|a=0)\leq \frac{\mu }{\mu +(1-\alpha )(1-\mu )} 
\]%
where $p_{01}=p_{00}=p_{11}=1$ and $p_{10}=0$ attain this upper bound. $%
\blacksquare $\bigskip

\noindent \textbf{Missing step in the proof of Proposition \ref%
{prop_opportunity}}

\noindent \textit{Let }$R^{o}:a\rightarrow y\leftarrow x_{2}.$\ Our
objective is to show that%
\begin{eqnarray*}
p_{R^{o}}(y &=&1|a=1)\leq 1-\alpha (1-\mu ) \\
p_{R^{o}}(y &=&1|a=0)\leq 1-(1-\alpha )(1-\mu )
\end{eqnarray*}%
in the $\delta \rightarrow 0$ limit. To derive these upper bounds, note
first that%
\[
p_{R^{o}}(y=1|a)=\sum_{x_{2}=0,1}p(x_{2})p(y=1|a,x_{2}) 
\]%
Denote $p_{ay}\equiv p(x_{2}|a,y).$ Then $p_{R^{o}}(y=1|a=1)$ is equal to%
\begin{eqnarray*}
&&\frac{{\small [\alpha \mu p}_{11}{\small +\alpha (1-\mu )p}_{10}{\small %
+(1-\alpha )\mu p}_{01}{\small +(1-\alpha )(1-\mu )p}_{00}{\small ]}\mu
\alpha p_{11}}{\alpha \lbrack \mu p_{11}+(1-\mu )p_{10}]}+\medskip \\
&&\frac{{\small [\alpha \mu (1-p}_{11}{\small )+\alpha (1-\mu )(1-p}_{10}%
{\small )+(1-\alpha )\mu (1-p}_{01}{\small )+(1-\alpha )(1-\mu )(1-p}_{00}%
{\small )]}\mu \alpha (1-p_{11})}{\alpha \lbrack \mu (1-p_{11})+(1-\mu
)(1-p_{10})]}
\end{eqnarray*}%
\medskip which simplifies into%
\begin{equation}
\lbrack 1+(\frac{1-\alpha }{\alpha })(\frac{\mu p_{01}+(1-\mu )p_{00}}{\mu
p_{11}+(1-\mu )p_{10}})]\mu \alpha p_{11}+[1+(\frac{1-\alpha }{\alpha })(%
\frac{\mu (1-p_{01})+(1-\mu )(1-p_{00})}{\mu (1-p_{11})+(1-\mu )(1-p_{10})}%
)]\mu \alpha (1-p_{11})  \label{pR(y|a) imp}
\end{equation}%
Note that this expression is a convex combination of two expressions,%
\begin{equation}
\frac{\mu p_{01}+(1-\mu )p_{00}}{\mu p_{11}+(1-\mu )p_{10}}  \label{A}
\end{equation}%
and%
\begin{equation}
\frac{\mu (1-p_{01})+(1-\mu )(1-p_{00})}{\mu (1-p_{11})+(1-\mu )(1-p_{10})}
\label{B}
\end{equation}%
Suppose (\ref{A}) is greater or equal to (\ref{B}). Then (\ref{pR(y|a) imp})
attains a maximum only if $p_{11}=1.$ Given this, (\ref{A}) attains a
maximum at $p_{01}=p_{00}=1$ and $p_{10}=0.$ Plugging these values into (\ref%
{pR(y|a) imp}) gives%
\[
p_{R^{o}}(y=1|a=1)=1-\alpha (1-\mu ) 
\]%
and (\ref{A}) is greater than (\ref{B}).

By analogous arguments,%
\[
p_{R^{o}}(y=1|a=0)\leq 1-(1-\alpha )(1-\mu ) 
\]%
and $p_{01}=p_{11}=p_{10}=1,p_{00}=0$ attain this upper bound. $\blacksquare 
$

\end{document}